\documentclass[%
reprint,
showpacs,
showkeys,
preprintnumbers,
amsmath,amssymb,aps,prb]{revtex4-1}
\usepackage{graphics}
\usepackage{epsfig}
\usepackage{dcolumn}
\usepackage{color}
\usepackage{bm}
\usepackage{subfigure}
\usepackage{hyperref}
\usepackage[mathlines]{lineno}

\begin{document}


\title{Thin films of a three-dimensional topological insulator in a strong magnetic field:\\
a microscopic study}

\author{A.~Pertsova$^1$, C.~M.~Canali$^1$ and A.~H.~MacDonald$^2$}
\affiliation{$^1$Department of Physics and Electrical Engineering,
Linn{\ae}us University, 391 82 Kalmar, Sweden\\
$^2$Department of Physics, University of Texas at Austin, TX 78712, USA}

\date{today}

\begin{abstract}
The response of thin films of Bi$_2$Se$_3$ 
to a strong perpendicular magnetic field is investigated 
 by performing magnetic bandstructure calculations for a realistic multi-band tight-binding model.  
 Several crucial features of Landau quantization in a 
realistic three-dimensional topological insulator are revealed. 
 The $n=0$ Landau level is absent in ultra-thin 
 films, in agreement with experiment. 
 In films with a crossover thickness of five quintuple layers, there is    
 a signature of the $n=0$ level, whose overall trend as a 
function of magnetic field matches the established 
 low-energy effective-model result. 
 Importantly, we find a field-dependent splitting and a strong spin-polarization of the 
$n=0$ level which can be measured experimentally at reasonable field strengths. Our calculations 
	 show  mixing between the surface and bulk Landau levels 
	 which causes the character of levels to evolve with magnetic field.  
\end{abstract}


\maketitle
\section{Introduction}

The peculiar structure of the Landau levels (LLs) present in three-dimensional 
(3D) topological insulators (TIs)~\cite{Hasan,XLQi,Hsieh2009} 
in a strong magnetic field is a characteristic signature  of 
their Dirac surface states~\cite{Cheng,Hanaguri,Okada}. 
In the simplest low-energy 
effective models, a field-independent $n=0$ level emerges at the surface-state 
band-crossing energy that is analogous to the $n=0$ level of another Dirac material 
-- graphene~\cite{Martin}, 
 suggesting strong similarities in the magnetic-field  response of these two systems. 
However, a number of important features have been observed recently 
in thin films of 
binary-chalcogenide 3D TIs 
which indicate 
that the conventional picture of Landau quantization may not be fully applicable to these materials. 
The most notable features include 
(\textit{i}) deviations from the square-root dependence of LL energies on magnetic field and LL index  
 which applies when linear dispersion is present over a wide energy regime~\cite{Cheng}, 
(\textit{ii}) asymmetry of the LL spectrum with 
respect to the Dirac point~\cite{Cheng,Hanaguri}, and (\textit{iii}) 
 finite-thickness effects, in particular the predicted splitting of the $n=0$ level 
 due to inter-surface coupling~\cite{Yang,Zyuzin,Tahir} 
 and its absence in ultra-thin films~\cite{Jiang}.  
 LL positions are readily measured in scanning probe studies and, since they depend 
 both on the zero-field energy bands and on the 
momentum-dependence of zero-field wavefunctions, 
are they a sensitive probe of the electronic structure.

In this work we study the electronic properties of 
thin films of Bi$_2$Se$_3$ 
in the presence of a strong quantizing 
magnetic field, using a microscopic approach which captures 
the complex electronic structure and the inter-surface hybridization of the realistic material. 
We perform magnetic bandstructure calculations~\cite{Brown1,Brown2,Graf} using   
 a multi-band tight-binding (TB) model for Bi$_2$Se$_3$~\cite{Kobayashi, NJP}. 
 We consider slabs of 1 to 6 quintuple layers (QLs) 
 in magnetic fields of the order of 10~T  
(for the smallest thickness) and larger. We find that the $n=0$ LL is absent in slabs with 
thicknesses below 5QLs but starts to emerge at this crossover thickness, in agreement with 
  experiments on a similar system (Sb$_2$Te$_3$)~\cite{Jiang}. Importantly, the energy gap 
  due to inter-surface coupling, which is found at the Dirac point of 3D TI 
	thin films at zero field~\cite{YZhang}, 
persists at finite magnetic fields. 

This finding is partly consistent with 
 recent theoretical studies, which investigated the 
 effect of finite thickness on the LL spectrum either by 
 introducing an \textit{ad hoc} hybridization gap into the Dirac Hamiltonian of 
the surface states~\cite{Zyuzin,Tahir} or by using a minimal TB 
 model~\cite{Yang} for a single-Dirac-cone family of 3D TIs.  
 However, the are crucial differences between the results obtained 
with our microscopic 
approach and with effective models. The hybridization 
 gap at finite magnetic fields emerges naturally in our electronic structure calculations for finite slabs.  
  As a result, the degeneracy of the $n=0$ level is lifted.   
   For moderate field strengths, the gap increases approximately linearly with the field, 
	with the dependence becoming weak with increasing thickness. 
	
  For 5QLs, the field-dependence of one of the two components of 
     the  $n=0$ LL is in good agreement with  
  the analytical expression derived by Liu \textit{et al.}~\cite{CXLiu} 
	using a four-band effective  Hamiltonian~\cite{Zhang2009}. 
	The other component, which 
   is energetically closer to the valence band, deviates further from the analytical curve 
   for increasing  magnetic fields, 
    with its wavefunction becoming progressively bulk-like.  
    In the limit of small fields, the splitting approaches the value 
    of the zero-field hybridization gap. 
 The LLs of the bulk and surface states, which can not be easily disentangled experimentally, 
 are identified based on the spatial character of their 
 wavefunctions. 
The splitting of the $n=0$ LL 
 can be detected by scanning tunneling spectroscopy (STS) experiments and 
can be used as a probe of pertinent electronic structure features. 

The paper is organized as follows. In Sec.~\ref{model} we describe the 
 TB model and discuss the details of magnetic bandstructure calculations. 
 The numerical results, namely the calculated LL spectra and the analysis of 
 the hybridization gap in the presence of applied magnetic 
 field for 1-6 QLs of Bi$_2$Se$_3$, are presented in Sec.~\ref{results}. 
  We also briefly outline possible implications of our findings for measurable 
  electronic properties of 3D TI thin films in magnetic field.
 Finally, we draw some conclusions.

\section{Model}\label{model}

The electronic structure of a Bi$_2$Se$_3$ slab  
 is described by a \textit{sp}$^3$ TB model with parameters  
obtained by fitting to DFT bandstructures~\cite{Kobayashi,NJP}. 
 The applied magnetic field is introduced via 
the Peierls substitution~\cite{Peierls, Hofstadter}. The Hamiltonian of the system reads

\begin{eqnarray}\label{eq:1} \hat{H}(\bf{k}) & = 
& \sum_{\substack{{\scriptscriptstyle ii^{\prime},\sigma}\\
{\scriptscriptstyle\alpha\alpha^{\prime}}}} \gamma_{\scriptsize ii^{\prime}}^{\scriptsize
\alpha\alpha^{\prime}}\,e^{i{\bf k}\cdot{\bf r}_{ii^{\prime}}}\,
e^{-\frac{i e}{\hbar c}\int_{i}^{i^{\prime}}{\bf A}\cdot d {\bf l}}\,
\hat{c}^{\sigma\dagger}_{i\alpha}
\,\hat{c}^{\sigma}_{i^{\prime}\alpha^{\prime}}\\\nonumber & + &
\sum_{\substack{{\scriptscriptstyle i,\sigma\sigma^{\prime}}\\{\scriptscriptstyle\alpha\alpha^{\prime}}}}
\lambda_{\scriptsize i} \left\langle i,\alpha,\sigma \right|
\hat{\vec{l}}\cdot\hat{\vec{s}}\left| i,\alpha^{\prime},\sigma^{\prime}
\right\rangle \,\hat{c}^{\sigma\dagger}_{i\alpha}
\,\hat{c}^{\sigma^{\prime}}_{i\alpha^{\prime}}\:, 
\end{eqnarray} 
where  $\mathbf{k}$ is the reciprocal-lattice vector, 
$i(i^{\prime})$ is the atomic index, $\alpha(\alpha^{\prime})$ labels atomic
orbitals, and $\sigma(\sigma^{\prime})$ denotes the spin. Here $i$ 
runs over all atoms in the magnetic unit cell (see definition below), while 
$i^{\prime}\ne i$ runs over all neighbors of atom $i$, including atoms in the adjacent cells; 
 $\mathbf{r}_{i i^{\prime}}$ is the distance between atoms $i$ and $i^{\prime}$ 
($\mathbf{r}_{i i^{\prime}}$=$0$ for $i$=$i^{\prime}$). 
 $\gamma_{\scriptsize i i^{\prime}}^{\scriptsize \alpha\alpha^{\prime}}$ are the Slater-Koster 
parameters and $\hat{c}_{i\alpha}^{\sigma\dagger}$ ($\hat{c}_{i\alpha}^{\sigma}$) 
is the creation (annihilation) operator for an electron with spin $\sigma$ at the atomic orbital $\alpha$ of site $i$.
We include the hopping between nearest neighbors in the $x$--$y$ 
plane and  next-nearest neighbors along the $z$--direction. 
The second term in Eq.~(\ref{eq:1}) is the intra-atomic spin-orbit interaction, 
where $\left|i,\alpha,\sigma\right\rangle$ are spin-
and orbital-resolved atomic orbitals, $\hat{\vec{l}}$ is the orbital angular
momentum operator and $\hat{\vec{s}}$ is the spin operator;
$\lambda_{\scriptsize i}$ is the SO strength. 

The factors $e^{i\theta_{i i^{\prime}}}$, with $\theta_{ii^{\prime}}$=$
-\frac{e}{\hbar c}\int_{i}^{i^{\prime}} {\bf A} \cdot d {\bf l}$ ($\theta_{ii}$=$0$), 
which multiply the hopping matrix elements in Eq.~(\ref{eq:1}), 
are the Peierls phase factors.  
 $\bf{A}$ is the magnetic vector potential and $\bf{l}$ is  
a straight path connecting the lattice sites $i$ and $i^{\prime}$. 
A uniform magnetic field is applied perpendicular to the surface of the slab, 
${\bf B}=B\hat{z}$, and we use the Landau gauge, ${\bf A}=(0,B\hat{x},0)$. 
We focus on the orbital contribution of the magnetic field, which is 
expected to be the most crucial one for Landau quantization, at least for magnetic fields in 
 the experimental range. Therefore we do not consider the Zeeman term. We are concerned 
 with the effect of the hybridization between the opposite 
surfaces, and between the surfaces and the valence band bulk, 
 on the LL spectrum of Bi$_2$Se$_3$ thin films.  
 
\begin{figure}[ht!]
\begin{center}\includegraphics[width=0.98\linewidth,clip=true]{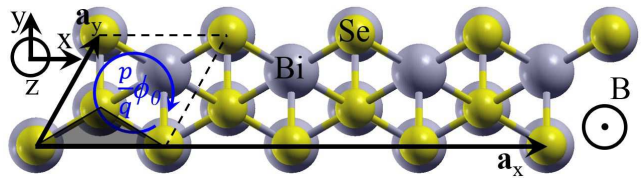}\end{center}
\caption{Top view ($x$--$y$ plane) of the magnetic unit cell ($q=4$) of 5QLs 
 of Bi$_2$Se$_3$. ${\bf a}_{x(y)}$ are the 2D lattice vectors.  
Magnetic field is along the $z$-axis. Dashed lines show the 
 crystal unit cell. Shaded triangle marks the elementary 2D placket.} 
\label{cryst}
\end{figure}

It is known that for electrons, subject to both magnetic 
field and a periodic potential, the electron wavefunction can not satisfy the 
periodic boundary conditions (the Hamiltonian in Eq.~(\ref{eq:1}) 
does not commute with the operator of translations). 
 However, one can introduce the new primitive translation vectors, 
 which define the \textit{magnetic unit cell}, 
 provided that the magnetic flux $\phi$ threading a crystal unit cell 
is a rational multiple of the magnetic flux quantum $\phi_0=hc/e$, i.e.  
 $\phi=p\phi_0/q$, where $p$ and $q$ are mutually prime integers~\cite{Brown1}. 
 The magnetic unit cell, carrying 
 a magnetic flux $p\phi_0$, is $q$ times larger then the 
crystal unit cell (see Fig.~\ref{cryst}). The corresponding magnetic Brillouin 
 zone is $q$ times smaller than the original one.  
 This \textit{magnetic periodic boundary condition} (MPBC) requirement  
guarantees that the electron wavefunction only accumulates 
 a phase $2\pi p$ when one moves along the edges of 
the magnetic unit cell. It follows from the MPBC that the value 
of the magnetic field is determined by the size of the magnetic unit cell, 
i.e. $B=p\,\phi_0/S_m=p\,\phi_0/q\,S_0$, where $S_m$ ($S_0$)   
is the area of the magnetic (crystal) unit cell in the 
 plane perpendicular to the field. This is a notorious 
  numerical limitation of magnetic bandstructure calculations.

The magnetic unit cell of a Bi$_2$Se$_3$ slab is built by replicating the slab unit cell $q$ times 
along the $x$-axis (Fig.~\ref{cryst}). Its size 
 grows as $q\times(5N_{\mathrm{QL}})$, where $N_{\mathrm{QL}}$ is the number of QLs. 
 With this constraint, we were able to 
reach minimum field strengths of 
 $\sim$8~Tesla for 1QL and 
$\sim 45$~Tesla for 5QLs. In order to predict  
 the behavior at smaller fields, we either use numerical fitting 
 or, when it is appropriate, interpolate the results 
  of our calculations between $B=0$ 
and the smallest field accessible 
  for a given thickness. We employ the Lanczos method of diagonalization at each $k$-point 
 in the magnetic Brillouin zone, and focus on a small 
energy window around the Dirac point to reduce the computational load. 
 
As one varies the parameter $p/q$, 
 a non-trivial fractal pattern in the electronic spectrum, 
known as the Hofstadter butterfly~\cite{Hofstadter}, 
emerges. First predicted for 
 electrons on a square two-dimensional (2D) lattice,  
  the pattern has been obtained for other 2D lattices  
  (honeycomb~\cite{Ramal}, triangular~\cite{triag}) 
  and a generalization to 
 the 3D case has been demonstrated~\cite{Hofstadter3D}. In this work we calculate  
 the Hofstadter butterfly for a slab of Bi$_2$Se$_3$ with varying thickness. 
By focusing on the low-field (low-flux) region of the 
 Hofstadter spectrum, which is typically the region probed in experiment,  
we will show the emergence of well-resolved LLs. 

\section{Numerical results}\label{results}
 
\begin{figure}[ht!]
\begin{center}\includegraphics[width=0.98\linewidth,clip=true]{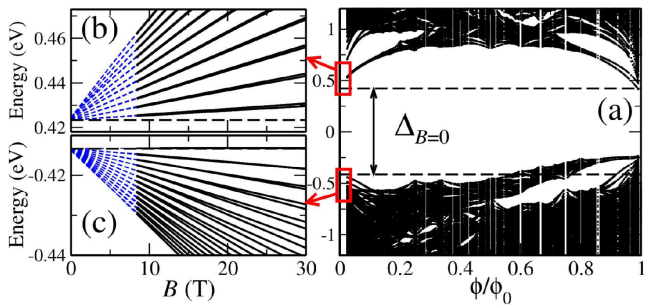}\end{center}
\caption{(a) Hofstadter spectrum of 1QL of Bi$_2$Se$_3$. 
Positive (b) and negative (c) branches of the LL spectrum (the lowest 24 levels are shown for each branch). 
Thin dashed lines show the results of numerical fitting for $B<8.3$~T. 
 Horizontal lines mark the zero-field energy gap.} 
\label{1QL}
\end{figure}

We start with the calculation 
 of LLs in ultra-thin films of Bi$_2$Se$_3$. 
 Figure~\ref{1QL}(a) shows the 
 Hofstadter spectrum of a 1QL-thick slab. 
 Only the results for $\phi/\phi_0\in\left[0,1\right ]$ are shown for better visibility. 
 As in the case of a one-band triangular lattice model~\cite{triag}, 
the spectrum  is not symmetric with respect to $\phi/\phi_0=1/2$ since 
the elementary placket, i.e. 
the smallest loop in the $x$--$y$ plane pierced by magnetic flux, is a fraction of 
the unit cell (Fig.~\ref{cryst}). At $B=0$ there is a large energy gap,  
$\Delta_{B=0}$=$0.84$~eV, due to inter-surface hybridization. 
This gap persists at finite magnetic fields. The field-dependence of 
the gap is quite complex, especially for large values of $\phi/\phi_0$, where  
 the interplay between the periodic lattice potential and the quantizing magnetic field is strong. 
 However, in the regime  
 where the  broadening of the LLs is sufficiently weak so that few lowest 
 levels can  be resolved ($\phi/\phi_0\lesssim 0.1$), the gap increases with  magnetic field. 
 
We now focus on two regions of the spectrum, marked by red boxes in Fig.~\ref{1QL}(a), 
corresponding to small fields ($B\le 30$~T) and  
 energies around the zero-field gap. 
 There are two distinct 
branches of the LL spectrum, positive and negative 
[Fig.~\ref{1QL}(b) and (c), respectively], with well-resolved levels. For the positive branch, 
the LLs come in pairs, originating from doubly generate states 
at $B=0$. The levels disperse almost linearly with magnetic field, 
in striking similarity with an ordinary 2D electron gas. 
A similar pattern is found for the negative branch, with 
the exception of the top LL, which depends weakly on $B$ in this range 
 but starts to deflect downwards for $B>40$~T. 
 
\begin{figure}[ht!]
\begin{center}\includegraphics[width=0.98\linewidth,clip=true]{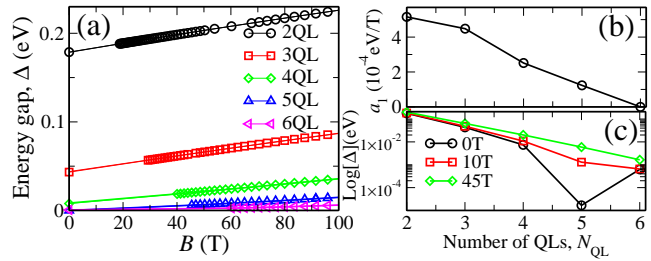}\end{center}
\caption{(a) Hybridization gap as a function of magnetic field (including $B=0$) for 2-6 QLs of Bi$_2$Se$_3$. 
Symbols show the data points. Solid lines are numerical interpolations in the low-field region. 
 Coefficient of the linear term in the field-dependence of the gap (b)  and   
 the logarithm of the gap (c) versus thickness for $B=0$, $10$ and $45$~T. 
} 
\label{2-6QLs}
\end{figure}

In the same way, we calculate the Hofstadter spectra and the LLs for 2-6 QLs. 
 The results are summarized in Fig.~\ref{2-6QLs}, where we analyze the hybridization gap 
 as a function of magnetic field and slab thickness. 
Note that for $B=10$~T we perform numerical interpolations  
 for all thicknesses except 1QL. For  $B=45$~T an interpolation is only required for the 
largest thickness of 6QLs. 
Figure~\ref{2-6QLs}(a) shows that the 
gap increases with magnetic field 
 for all slabs considered. The dependence is predominantly linear. 
 By using numerical fitting, we determine numerically the linear (dominant) coefficient 
 $a_1$ in the field-dependence 
 and plot it versus the thickness in Fig.~\ref{2-6QLs}(b). As one can see, 
 the field-dependence becomes weaker for thicker slabs. At the same time,  
   the gap decreases exponentially with the thickness, as shown in Fig.~\ref{2-6QLs}(c).  
 At $B=0$ this result is well-established. However, here  
   we explicitly demonstrate the exponential decay at finite fields. 
 As the thickness increases beyond 5QLs, the gap becomes exponentially small and field-independent. 
Hence, in the limit of an infinitely-thick slab the expected LL structure, 
with a doubly-degenerate field-independent $n=0$ level, is recovered. 
 
The existence of the hybridization gap at finite magnetic fields was investigated in Ref.~\onlinecite{Yang} 
using the low-energy model of Refs.~\onlinecite{YZhang,CXLiu}. Within this approach the particle-hole symmetry 
is imposed from start, and therefore the model is unable to capture the asymmetry of the LL spectrum 
and the strong hybridization with the valence band, which is significant in Bi$_2$Se$_3$. Our approach is free 
from these limitations and  provides  qualitatively and quantitatively 
accurate description of these features 
that are crucial in thin films.

In Ref.~\onlinecite{Yang} the gap was found to depend weakly on magnetic field,   
  oscillating as a function of the thickness, in analogy with the zero-field gap 
 found in other continuum models~\cite{CXLiu,Linder,Lu2010}. We find that at $B=0$ 
 the gap exhibits non-monotonic behavior at certain thicknesses, which resembles the oscillations in   
 these continuum models. 
The non-monotonicity is also present at $B\ne 0$, but is smoothed out 
at large fields [Fig.~\ref{2-6QLs}(c)]. The analytical expression for the gap in Ref.~\onlinecite{Yang} 
contains a term linear in $B$ (multiplied by an oscillatory and exponentially decaying function of thickness). 
  This is consistent 
  with our calculations but only for moderate field strengths:
   at large fields the field-dependence in our model is highly nonlinear [Fig.~\ref{1QL}(a)], 
	 in contrast to the result of the continuum model.

The field-dependence of LLs that we find in ultra-thin films 
 at low magnetic fields (Fig.~\ref{1QL}) 
 is clearly distinct from the square-root behavior, expected from  the  
 Dirac-Hamiltonian description of the surface states. 
 Together with the absence of the $n=0$ level and a sizable hybridization gap below 5QLs (Fig.~\ref{2-6QLs}), 
   these results demonstrate that the Dirac fermion picture 
  does not capture the magnetic field response of 3D TI thin films 
	at moderate magnetic field strengths. 

\begin{figure}[ht!]
\begin{center}\includegraphics[width=0.99\linewidth,clip=true]{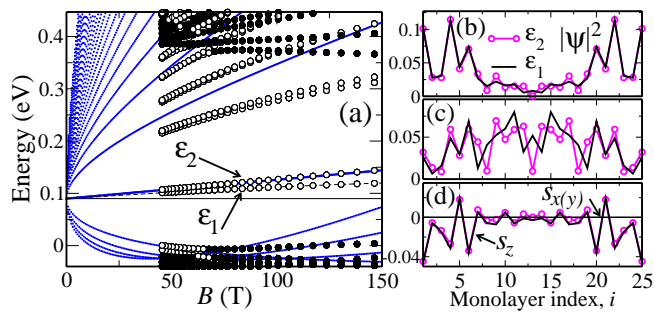}\end{center}
\caption{(a) LL spectrum of 5QLs of Bi$_2$Se$_3$. Open (filled) 
 symbols are for surface (bulk) states. Dashed lines are numerical interpolations 
for the $n=0$ level. Solid lines are the analytical surface LLs from Ref.~\onlinecite{CXLiu}. 
 Horizontal line in (a) marks the position of the Dirac point at $B=0$. 
Wavefunctions $|\psi|^2$ (b,c) and expectation values of the spins $s_{x(y,z)}$ (d) 
of the two components of the $n=0$ LL, marked as $\varepsilon_1$ and $\varepsilon_2$ in (a), 
plotted versus distance along the width of the slab. 
In (b) and (d) $B=45$~T, in (c) $B$ is ten times larger.
} 
\label{5QL}
\end{figure}

For typical magnetic field strengths and 
disorder strength, the 5QL thickness marks a crossover to the range beyond 
which the topologically protected bulk TI surface state is robustly manifested~\cite{YZhang}. 
This suggests that a nearly degenerate 
 $n=0$ LL is expected to appear at this thickness at finite magnetic fields. 
The calculated LL spectrum for 5QLs 
is shown in Fig.~\ref{5QL}(a).  
We compare our results with the analytical expression for the LLs 
 of the surface states, obtained in Ref.~\onlinecite{CXLiu}, 
 using a low-energy effective Hamiltonian under 
   the assumption of decoupled bulk and surface states. 
 
 By setting the Zeeman term to zero in the analytical 
formula of Liu \textit{et al.}~\cite{CXLiu}, 
 we find that the energy of the $n=0$ level is given by $E_0=\tilde{C}_0+e \tilde{C}_2 B/\hbar$,
   where $\tilde{C}_{0(1)}$ are the parameters of the model. The term $\propto\tilde{C}_2 B$ 
   can be traced back to the non-linear (quadratic) term $\tilde{C}_2 k^2$ in 
   the Hamiltonian of the surface states as a function of momentum $k$, 
	characteristic of the  complex electronic structure of the material. 
   Although at low fields the $n=0$ level is nearly constant, the field-dependence becomes significant already 
   at $B\sim20$~T. 
   
   In our calculations at finite fields, two 
   distinct levels with a quasi-linear dispersion   
 emerge close to the original Dirac point. We interpret this pair of levels as the two 
  components of $n=0$ LL, split by inter-surface coupling. This splitting 
  is not captured by the effective model of Ref.~\onlinecite{CXLiu} 
	 for the surface states of a semi-infinite system.    
  However, the field-dependence  
	of the highest of these two levels matches 
	  remarkably well the analytical curve (the position of the Dirac point 
	   in the analytical expression 
	  has been adjusted to that found in our calculations, i.e. $\sim 0.09$~eV).  
    In the limit $B\rightarrow 0$, we expect the 
    numerical and analytical results to converge, as confirmed by 
     numerical interpolation (apart 
    from a small but non-negligible gap at zero field). 
    For the higher-index LLs, we find significant 
		deviations from the analytical results. 
    These features are due to 
  the finite thickness of the sample 
and to higher-order non-linearities in the electronic structure that are captured by our model.
    
  The two components of the $n=0$ level are the true 
surface states, as one can see from the real-space distribution of their wavefunctions  
    along the slab [Fig.~\ref{5QL}(b)]. 
		In fact, their wavefunctions are almost indistinguishable from  the ones calculated at zero field. 
    The surface-character is preserved in fields as large as $100$~T. The levels have 
   the same spin polarization and are nearly fully spin-polarized in the 
   direction opposite to magnetic field [Fig.~\ref{5QL}(d)]. This finding   
   is consistent with the analytical prediction of Ref.~\onlinecite{Yang}. 
    The splitting between the two $n=0$ levels increases with magnetic field. 
For 5QLs, we find a splitting of 6~meV at $45$~T and we estimate the 
splitting to be $3$~meV at $20$~T.

Our calculations 
capture the strong asymmetry of the LL spectrum with respect to the Dirac point observed 
in experiments~\cite{Cheng,Hanaguri} and the mixing between bulk and surface states. 
 The asymmetry is mainly due to the fact that the Dirac point is close  
 to the valence band. As a result, the negative branch has only few 
well-resolved LLs, which merge with the bulk states with increasing magnetic field. 
 On the contrary, the positive branch contains many levels which preserve their 
surface character at large fields. 
Based on the real-space analysis of wavefunctions, 
 the states roughly in the energy window $[-0.02;0.4]$~eV are surface states,
 while the ones outside this window are bulk states. Importantly, 
 the character of the states does not remain constant, and  
 some of the states evolve from surface- to bulk-like as the field increases 
[see Fig.~\ref{5QL}(a)]. 

This behavior is due to proximity of the bulk and  
can be understood using the following semi-classical argument. 
With increasing magnetic field, the radius of the $n$-th Landau orbit in momentum space, 
expressed as $k_n=\sqrt{2|n|}/l_B$  
  with $l_B=\sqrt{\hbar/|e|B}$ being the magnetic length~\cite{Hanaguri}, increases. Therefore, the LLs  
  start to involve states at $k$-points further away 
from $\Gamma$. In the bandstructure of a Bi$_2$Se$_3$ thin film, at $k\sim k_n$ 
the energy bands can be quite different from the linearly dispersed Dirac 
states, especially below the Dirac point, i.e. 
 close to the valence band. Indeed, 
for 5QLs at $k\approx k_n$, 
with $k_n\approx 0.3$~\AA$^{-1}$ for $B\approx 50$~T and $|n|=1$, 
the bulk contribution to the 
 energy bands just below the Dirac point becomes dominant~\cite{NJP}. This also 
explains why at large fields the lower component of the $n=0$ LL deflects towards 
the valence band, while the upper one still follows the analytical curve. Since the lower component is energetically 
closer to the valence band, it is more affected by the valence-band states and  
at very large fields its wavefunction becomes more bulk-like [see Fig.~\ref{5QL}(c)].
               
LLs manifest as peaks in the electronic 
density of states that can be detected in 
 tunneling spectra measured by scanning tunneling spectroscopy (STS)~\cite{Cheng,Hanaguri}. 
  The energy of the $n=0$ LL indicates the position of the Dirac point within the energy bands.  
  A scaling analysis of LLs can be used to accurately determine the energy dispersion 
 in 3D TIs~\cite{Hanaguri}. 
  The splitting of the $n=0$ LL studied in this work is detectable by STS 
  at moderate magnetic fields. Hence, it can be used 
  as an alternative sensitive probe of the inter-surface hybridization, and  
  in particular of the hybridization gap. 
	
LLs are typically associated with the physics of the quantum Hall effect (QHE). 
  The structure of LLs is embodied in the 
  optical conductivity tensor calculated in the presence of a magnetic field, in the framework of linear response 
   theory. This yields an estimate of the Hall conductivity in the QHE regime 
   and allows, in general, the calculation of magneto-optical effects~\cite{Allan}.  
We expect the features of the LLs in 3D TI thin films  predicted in the present work to 
affect the structure and position of the Hall plateaus as function of the chemical 
potential~\cite{Tahir, Zyuzin}.

\section{Conclusions}

In conclusion, we presented a microscopic study of Landau quantization 
 in thin films of Bi$_2$Se$_3$. We find that 
the $n=0$ level is absent in ultra-thin films. For a thin film 
containing 5QLs, the degeneracy of the $n=0$ level is 
lifted due to 
hybridization between top and 
bottom surface states, with the two components being 
strongly spin-polarized. Since 
it is now possible 
to probe the properties 
of 3D TIs in 
strong magnetic fields ($\gtrsim 20$~T)~\cite{Analytis} and 
 to measure sub-meV gaps in thin films of 3D TIs~\cite{Kim}, 
these spin-polarized states with a splitting of few meV can be measured experimentally. 
The non-trivial structure of Landau levels, originating from the realistic bandstructure and 
 the inter-surface coupling, will affect the properties of 3D TI thin films in magnetic field, 
 in particular magneto-optical properties, the surface quantum Hall effect 
and the quantum anomalous Hall effect.

\begin{acknowledgements}

This work was supported by the Faculty of Natural Sciences at Linnaeus University and by the
Swedish Research Council under Grant Number: 621-2010-3761. 
AHM was supported by the Welch Foundation under 
Grant No. TBF1473 and by the DOE Division of Materials 
Sciences and Engineering under grant No. DE-FG03-02ER45958. 
Computational resources have
been provided by the Lunarc center for scientific and technical computing at 
Lund University. 
\end{acknowledgements}

\bibliography{pertsova}

\end{document}